\documentclass[aps,prl,twocolumn,amsfonts,showpacs]{revtex4} 
\usepackage{epsfig,amsopn}
\begin{document}

\title{Pattern formation inside bacteria: fluctuations due
to low copy number of proteins}  

\author{ Martin Howard$^{1,*}$  and Andrew D. Rutenberg$^{2}$ }

\affiliation{$^1$Instituut--Lorentz, Leiden University, PO Box 9506,
2300 RA Leiden, The Netherlands}

\affiliation{$^2$Department of Physics, Dalhousie University, Halifax,
Nova Scotia, Canada B3H 3J5}

\begin{abstract} 
We examine fluctuation effects due to the low copy number of proteins
involved in pattern--forming dynamics within a bacterium. We focus on
a stochastic model of the oscillating MinCDE protein system regulating 
accurate cell division in {\em E. coli}. We find that, for some
parameter regions, the protein concentrations are low enough that
fluctuations are {\em essential} for the generation of patterns. 
We also examine the role of fluctuations in constraining
protein concentration levels.  

\end{abstract}  

\pacs{87.17.Ee, 87.16.Ac, 05.40.-a}

\date{\today} \maketitle 

In recent years, dramatic experimental progress has been made in
resolving the subcellular localization of bacterial proteins.
Often the proteins form self--organized, spatially
inhomogeneous patterns that can involve coherent spatiotemporal
oscillations. These self-organized patterns are 
vital for accurate cell division
\cite{Raskin,Raskin1,Rothfield,Hu,Rowland,Fu,Hale}.
Understanding these patterns promises to reveal
new mechanisms for the generation of subcellular bacterial structure. 
In this letter, we examine, for the first time, the impact of
fluctuations on these
patterns, focusing on the oscillatory MinCDE system in {\em E. coli}
\cite{Raskin,Raskin1,Hu,Rowland,Fu,Rothfield,Hale}. 

Each {\em E. coli} cell divides roughly every hour, first replicating
its DNA into two nucleoids, and then dividing at midcell into
two daughter cells. If division is not targeted accurately to midcell
then DNA will not be distributed to both daughter cells,
resulting in unviable anucleate ``minicells''.  Division is initiated
by a polymeric ring of the protein FtsZ, which forms on the inner side
of the cytoplasmic membrane.  Precise positioning of the FtsZ
ring to midcell is controlled both by the inhibiting effect of the
nucleoids (``nucleoid occlusion''), and by the MinCDE
system of proteins \cite{Yu}. 
MinC inhibits the formation of the FtsZ ring, and is
recruited to the membrane by MinD. MinE is also recruited
to the membrane by MinD, where it forms a dynamical ring structure
that drives pole to pole oscillations of MinC and MinD, with a
period of about a minute. The oscillations lead to a
time--averaged midcell MinC concentration
minimum \cite{Hale}. As MinC inhibits
FtsZ ring formation, the FtsZ can only assemble near the cell
midplane. The self--organized protein
patterns are thus used by the cell to obtain positional
information without stationary  positional markers
\cite{Howard,MeinhardtBoer,Kruse}.

Very recently, the
physical principles behind the MinCDE protein patterns
have been explored through reaction--diffusion equations
\cite{Howard,MeinhardtBoer,Kruse}. The resulting equations represent
protein diffusion, both in the cytoplasm and along the membrane, and
also protein binding/unbinding from the cytoplasmic
membrane. The slow membrane diffusion used in these models,
relative to the cytoplasmic diffusion, results in 
Turing--like (Hopf) instabilities that spontaneously generate
oscillatory patterns, in good agreement with experiment
\cite{Howard,MeinhardtBoer,Kruse}. 

Bacterial proteins are, however, typically present in low numbers
within the cell. This induces large fluctuations \cite{Mcadams,Bialek}
which have not been considered in previous pattern--forming models
\cite{Howard,MeinhardtBoer,Kruse}. In {\em E. coli}, 
a recent assay put the copy numbers for MinD and MinE at 2000
and 1400 respectively \cite{Shih}. We have therefore investigated
the role of fluctuations in a discrete particle model of the
{\em E. coli} MinCDE system where each protein molecule is explicitly
tracked. This allows for a full analysis of the
fluctuations of both reactions and diffusion. Although
the effects of noise have been studied in
subcellular models without spatial dynamics \cite{Paulsson,Vilar}, 
and also in some spatially 
extended patterns \cite{Kessler,Falcke,Vilar1,Sancho,Z},  
the effects of fluctuations on {\em subcellular}
positional information, and the constraints on protein
concentrations due to fluctuations,
have not previously been studied. We find that:

\noindent (i) 
Even at surprisingly low concentrations, the noise does not destroy 
the oscillatory dynamics and indeed can be vital for generating patterns 
in regions of parameter space where the equivalent deterministic dynamics
decays away. A bacterium can thus exploit low copy number fluctuations
to produce stable, self--organized patterns. This 
result likely applies to {\em any} stochastic reaction--diffusion model of 
pattern formation involving sufficiently few protein copies
(i.e. thousands or fewer).

\noindent (ii) 
We find evidence that the cell employs sufficient copy numbers
of the MinCDE proteins to ensure reliable midcell division;
using substantially fewer copies degrades 
accurate positioning of the midcell MinCD minimum, using more does not 
lead to significantly improved accuracy. 

{\em Stochastic model.}
We begin by introducing the stochastic model for the MinCDE dynamics,
based on our deterministic model of Ref.~\cite{Howard}.
We employ a $1d$ discrete particle model, where the particles hop
between lattice sites and where the full
fluctuation effects are intrinsically included by discrete particles. 
As in earlier models
\cite{Howard,MeinhardtBoer,Kruse}, the MinC dynamics is omitted, since
it is known to closely follow the MinD dynamics.
The occupancy at site $i$ is $n_j^{\{i\}}$, with $j=\{D,d,E,e\}$
representing cytoplasmic MinD, membrane MinD, cytoplasmic MinE, and
membrane MinE, respectively. Each protein molecule is represented as a
particle which, at each timestep, may hop with equal probability
$\tilde D_j \Delta t/(\Delta x)^2$, where $\Delta t$ is the time
increment, to one of its neighboring sites at $x\to x\pm\Delta
x$ (except for the boundary sites at either end, where hard wall
boundary conditions are imposed). At site $i$, the following
reactions may occur, the first being for each $D$
particle, and then for each $d,E,e$ particle respectively:
\begin{eqnarray}
& & \hspace{5.65cm} {\rm Probability:} \nonumber \\
& & \hspace{-0.4cm}
n_D^{\{i\}}\to n_D^{\{i\}}-1,~n_d^{\{i\}}\to n_d^{\{i\}}+1
\hspace{0.5cm} 
\tilde \sigma_1 \Delta t/(1+\tilde \sigma_1'n_e^{\{i\}}) ~ , \nonumber \\
& & \hspace{-0.4cm}
n_D^{\{i\}}\to n_D^{\{i\}}+1,~n_d^{\{i\}}\to n_d^{\{i\}}-1
\hspace{0.5cm} 
\tilde \sigma_2 \, \Delta t \, n_e^{\{i\}} ~ ,
\nonumber \\
& & \hspace{-0.4cm}
n_E^{\{i\}}\to n_E^{\{i\}}-1,~n_e^{\{i\}}\to n_e^{\{i\}}+1 
\hspace{0.5cm}
\tilde \sigma_3 \, \Delta t \, n_D^{\{i\}} ~ ,
\nonumber \\
& & \hspace{-0.4cm}
n_E^{\{i\}}\to n_E^{\{i\}}+1,~n_e^{\{i\}}\to n_e^{\{i\}}-1 
\hspace{0.5cm}
\tilde \sigma_4\Delta t/(1+\tilde \sigma_4' n_D^{\{i\}}) ~ .
\nonumber
\end{eqnarray}
These reactions are the stochastic analogs of the
reaction processes used in our deterministic partial differential
equation model \cite{Howard}. The $\tilde \sigma_1$ term describes 
spontaneous 
membrane association of MinD, whereas the $\tilde \sigma_2$ term
describes ejection of MinD from the membrane by membrane--bound
MinE. Similarly, the $\tilde \sigma_4$ term describes spontaneous
membrane disassociation of MinE, whereas the $\tilde \sigma_3$
term describes recruitment of MinE to the membrane by cytoplasmic
MinD. 
The $\tilde \sigma_1'$, $\tilde \sigma_4'$
``suppression'' terms correspond to 
membrane MinE suppressing the binding of MinD to the membrane,
and to cytoplasmic MinD suppressing the unbinding of membrane 
MinE.

{\it ATP dynamics.} 
We next address the question of how ATP dynamics fits
into the model. Experimentally, MinD is an 
ATPase \cite{deBoer} and it is the MinD-ATP complex that 
binds to the membrane, whereas the release of MinD back into the
cytoplasm requires MinE--induced ATP hydrolysis \cite{Hu02}. Our
model assumes that, following ATP hydrolysis and release of MinD into
the cytoplasm, nucleotide exchange is sufficiently rapid to allow for
membrane reattachment of MinD-ATP almost immediately.  
As a result, we only model MinD-ATP in the cytoplasm ($n_D^{\{i\}}$)
and on the membrane ($n_d^{\{i\}}$). 
Like ``actin treadmilling''  
in eukaryotic cells, the ATP driven binding and
unbinding of MinD allows for a cyclic MinCDE pattern to be maintained
with only low levels of protein synthesis:
{\em this makes the pattern--forming dynamics extremely energy
efficient}. This recycling is also consistent 
with experiments where protein synthesis was blocked, but where 
the MinCDE oscillations were observed to continue unaffected
\cite{Raskin}. Consequently, the above model (similar to that of
Ref.~\cite{Kruse}) does not include protein synthesis or
degradation. These processes occur, but
on longer time scales than the relatively 
rapid MinCDE oscillations. 

{\em Simulations.}
In our stochastic model simulations, we use time
and spatial increments $\Delta x=0.02~\mu m$, $\Delta
t=2\times 10^{-5}~s$, so that $100$ lattice sites model a 
$2~\mu m$ bacterium.
We use: $\tilde D_D=0.28~\mu
m^2~s^{-1}$, $\tilde D_d=0.003 ~\mu m^2~s^{-1}$, $\tilde D_E=0.6~\mu
m^2~s^{-1}$, $\tilde D_e=0.006~\mu
m^2~s^{-1}$, $\tilde \sigma_1=20~s^{-1}$, and
$\tilde \sigma_4=0.8~s^{-1}$. Note that the membrane diffusion
constants are much smaller than those in the cytoplasm; this agrees
with recent data indicating that MinD may polymerize on the
membrane \cite{Hu02,polymer}.  
For the remaining variables of the model, we focus on
4 representative parameter sets shown in Table~I, where we define $N$
as the total number of MinD proteins, equal to the total number of
MinE proteins \cite{param}. However, 
we emphasize that our results for the oscillatory behavior observed 
below are typical for large regions of parameter space.
Initially, MinD and MinE particles are 
randomly distributed on the membrane and in the cytoplasm. Equal 
numbers of proteins are used since 
``wild--type'' oscillations are observed when both proteins are
equally expressed on plasmids (this is consistent, within experimental
uncertainties, with the earlier quoted MinDE assay \cite{Shih}).

{\em Fluctuation driven instability.}
Using the parameters in Table~I, we find that the
presence of noise is vital for the oscillations to persist.
This is shown in Fig.~1, where the ratio of the average MinD density
in the right--hand $30\%$ of the cell to that in the left--hand
$30\%$ of the cell is plotted as a function of time for the stochastic
model (at $N=200$ and $N=1500$) and for the deterministic model
\cite{Howard} with equivalent parameters (i.e. with the above reaction 
probabilities directly transformed into deterministic reaction rates). 
In both cases, for the deterministic model, the
protein concentrations rapidly decay away to the homogeneous steady
state (in agreement with linear stability analysis
\cite{Howard}), whereas regular oscillations continue for
the stochastic model. Hence the average behavior of the stochastic model
is clearly not describable using the naively equivalent deterministic 
model. To investigate this issue in more detail, we have examined
how steady--state/oscillation bifurcations in the deterministic 
model are altered in the stochastic model. As a representative 
example, at $N=1500$, using the above parameter set in the stochastic 
model, but with $\tilde D_d=0.001 ~\mu m^2~s^{-1}$, 
$\tilde D_e=0.003 ~\mu m^2~s^{-1}$, 
$\tilde\sigma_1'=0.2$ and varying $\tilde\sigma_4$, we find that the
transition from oscillatory to steady--state behavior is 
reduced by around $40\%$ from $\tilde\sigma_4=0.63~s^{-1}$ 
(deterministic) to $\tilde\sigma_4\approx 0.39~s^{-1}$ (stochastic), as 
determined using the end to end MinD ratio. 
The noise does smear out the transition somewhat in the 
stochastic model (the transition at $\tilde\sigma_4\approx 0.39~s^{-1}$ has 
width $\pm 0.05~s^{-1}$), but this effect is rather small. Hence this 
smearing out
cannot account for the large regions of parameter space where
oscillations occur in the stochastic, but not the deterministic, model.
Rather, we have a fluctuation driven instability, where the noise has 
shifted the location of the transition,
thereby promoting oscillations in large regions of parameter space
where it would be forbidden in the equivalent deterministic model (see
Fig.~1). {\em Cells can in principle exploit low copy 
number fluctuations to generate 
pattern--forming dynamics}. The oscillations continue down to very low 
concentrations ($N=200$), underlining the robustness of the
dynamics. 

\begin{table}
\begin{center}
\begin{tabular}{|c|c|c|c|c|}
\hline
$N$ & $\tilde \sigma_1'$ & $\tilde \sigma_2$ $(s^{-1})$ & 
$\tilde \sigma_3$ $(s^{-1})$ & $\tilde \sigma_4'$ \\ \hline
200 & 25.0 & 0.27 & 30.0 & 20.0 \\ 
400 & 2.0 & 0.135 & 15.0 & 10.0 \\
800 & 0.6 & 0.0675 & 7.5 & 5.0 \\
1500 & 0.25 & 0.036 & 4.0 & 2.7 \\
\hline
\end{tabular}
\end{center}
\caption{Reaction rate parameter values.}
\end{table}

The above examples show that the fluctuations are often
essential for pattern formation. However, it
also possible for the stochastic and deterministic models with
equivalent parameters \cite{Howard} both to generate
oscillations. Hence we cannot definitively conclude that
fluctuations are essential for the MinCDE oscillations.
Nevertheless, our analysis does show conclusively that cells can
exploit low copy number fluctuations for the
generation of dynamical subcellular structure. 

{\em Effect of fluctuations on the midcell MinD minimum.}
In Fig.~2, we plot the MinD and MinE concentration
profiles for $N=200$ and $N=1500$, showing their averages over $160$
and $110$ successive cycles, respectively, and also for $4$ individual 
data sets each, where each set is averaged over only an
{\em individual} oscillation cycle. For the long time data sets, we 
find that the midcell MinD concentration minimum (and a MinE 
concentration maximum) are still robustly reproduced even in the 
presence of noise. However, as can be seen from the data averaged over   
individual cycles, the fluctuations around this average can be very
large for small $N$. In Fig.~3 we show histograms of the position of the 
MinD concentration minimum, where each minimum is determined over a single
oscillation cycle (the use of a single cycle here is explained
below). For $N=1500$, the histogram 
is sharply peaked around the cell center at $1.0\pm 0.07~\mu m$ ($1$ 
standard deviation).
As expected, with decreasing $N$ the width increases: 
$0.09~\mu m$ at $N=800$, $0.16~\mu m$ at $N=400$, and 
$0.27~\mu m$ at $N=200$ \cite{param}.  
The width of the midcell localization is large, particularly at 
protein counts ($N=200$ and $N=400$) that are significantly below
those seen  naturally. Hence, using significantly fewer protein
copies degrades the accuracy of midcell division. 

Nucleoids, when present, also affect the positioning of the FtsZ ring 
through the poorly understood phenomenon of ``nucleoid occlusion''
\cite{Yu}, where FtsZ rings do not nucleate over nucleoids and
are restricted to either near the  midcell or at the cell poles.   
Segregated nucleoids (at the $1/4$ and
$3/4$ positions along the cell) will truncate the tails of the 
distributions shown in Fig.~3, further enhancing the accuracy of
midcell division (in agreement with experiment
\cite{sun1998,Yu}). 
In normal cells with nucleoids, it is particularly important that the
MinCDE system block polar FtsZ rings, since the nucleoids
themselves will inhibit FtsZ rings elsewhere away from midcell. Assuming
that FtsZ nucleation occurs at a single cycle MinD minimum (see below), 
then from Fig.~3 we see that $N=1500$ is a high enough
concentration to reduce the probability of polar division to 
considerably less than $0.01$ per oscillation cycle. Given that about $50$ 
complete oscillation cycles normally occur between successive 
divisions, we see 
that attaining this level of accuracy is 
important. Significantly lower 
concentrations than $N=1500$ will lead to an
unacceptable probability of polar division, while higher concentrations 
will lead to only marginally increased accuracy, but at the cost of 
manufacturing
many additional protein copies. From these simple arguments based on 
fluctuation effects, we see that {\em E. coli} may be using an optimal 
number of Min proteins, trading off midpoint precision against the
cost of protein synthesis. There will be other constraints on 
the protein copy numbers (e.g., sufficient MinC to
successfully inhibit off--center FtsZ ring formation), but fluctuations 
set useful bounds on the concentration levels. 

\begin{figure}
\centerline{\hbox{
\hspace{0.3cm}
\epsfysize=1.5in
\epsfbox{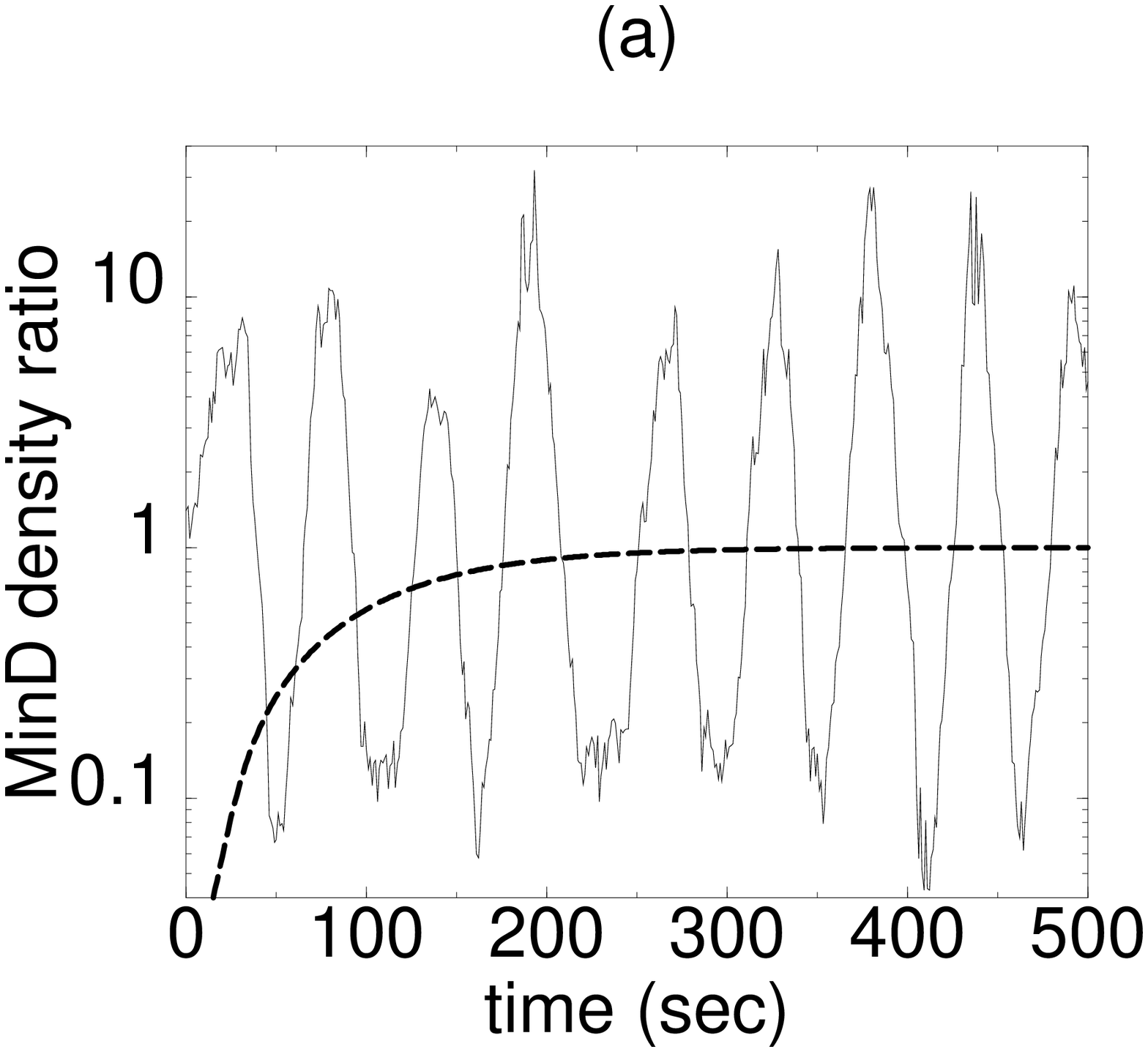}}
\hspace{.06in}
\epsfysize=1.5in
\epsfbox{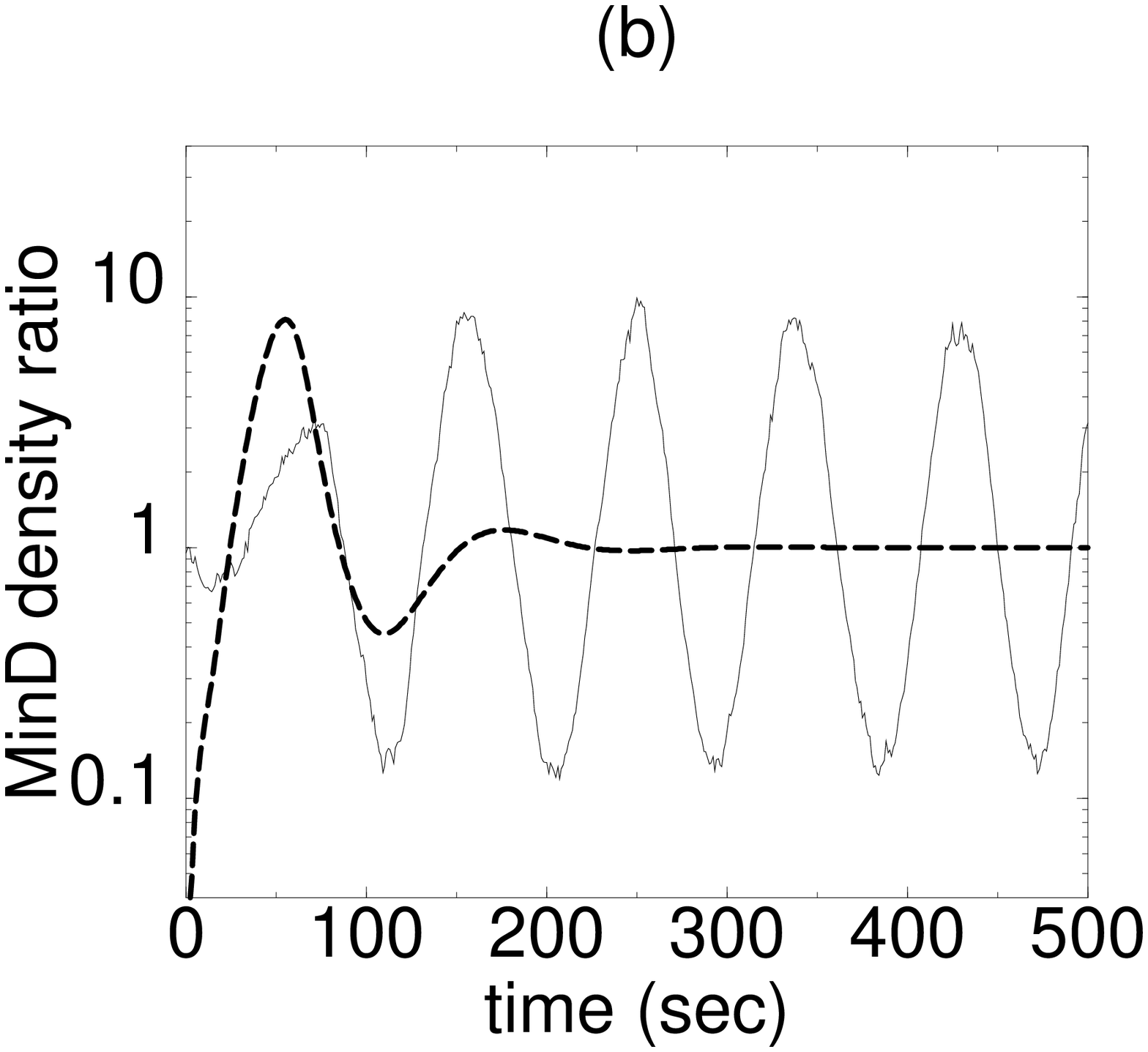}
}
\caption{
Ratio of average MinD density in right--hand $30 \% $ of the cell 
to that in left--hand $30 \% $. Full line: stochastic
dynamics with random initial conditions at (a) $N=200$ (b) $N=1500$;
dashed line: deterministic dynamics with equivalent parameters but
with inhomogeneous initial conditions. }
\label{stochvspde}
\end{figure}

\begin{figure}
\vbox{
\centerline{\hbox{
\hspace{0.2cm}
\epsfysize=1.5in
\epsfbox{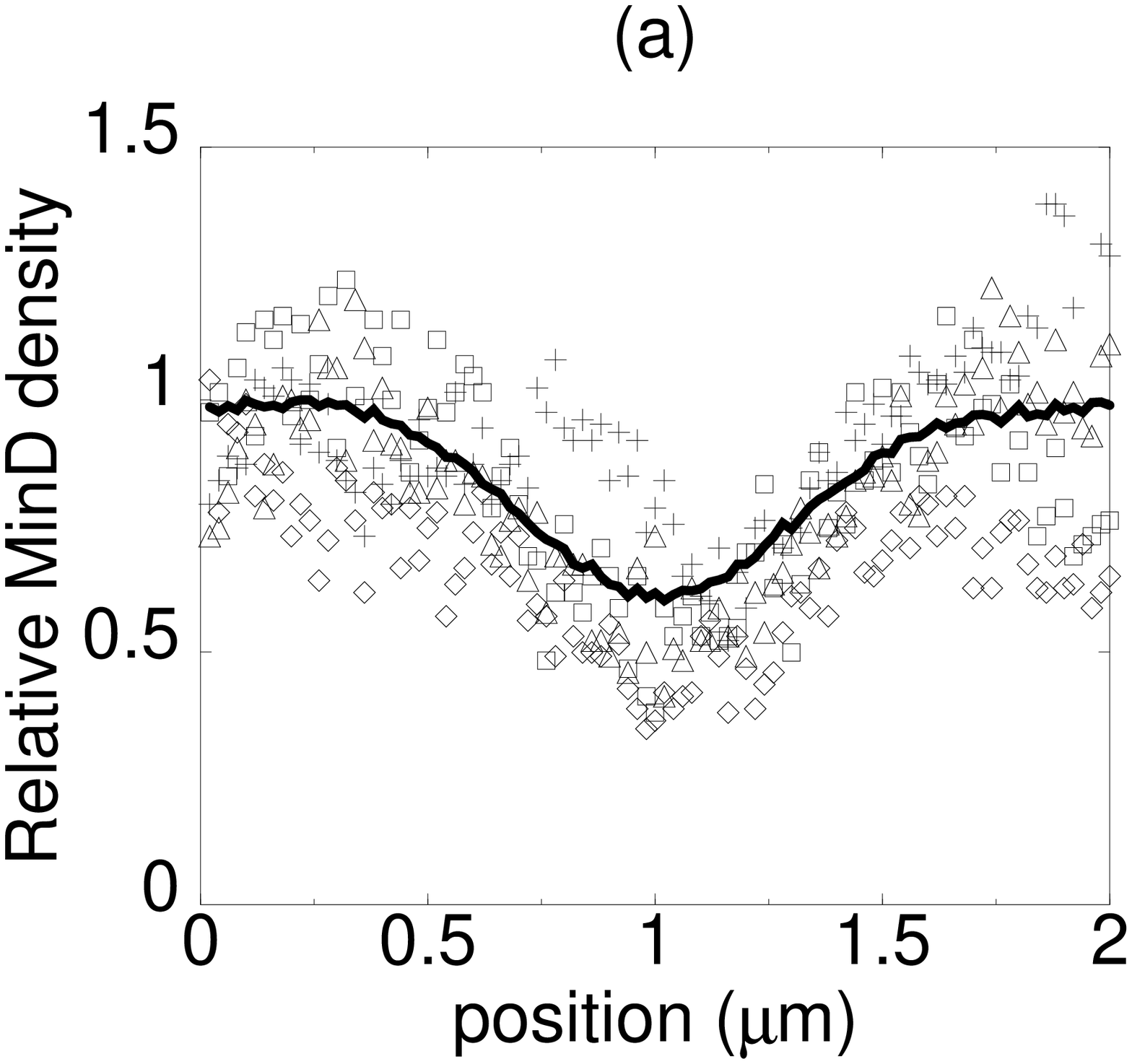}}
\hspace{.1in}
\epsfysize=1.5in
\epsfbox{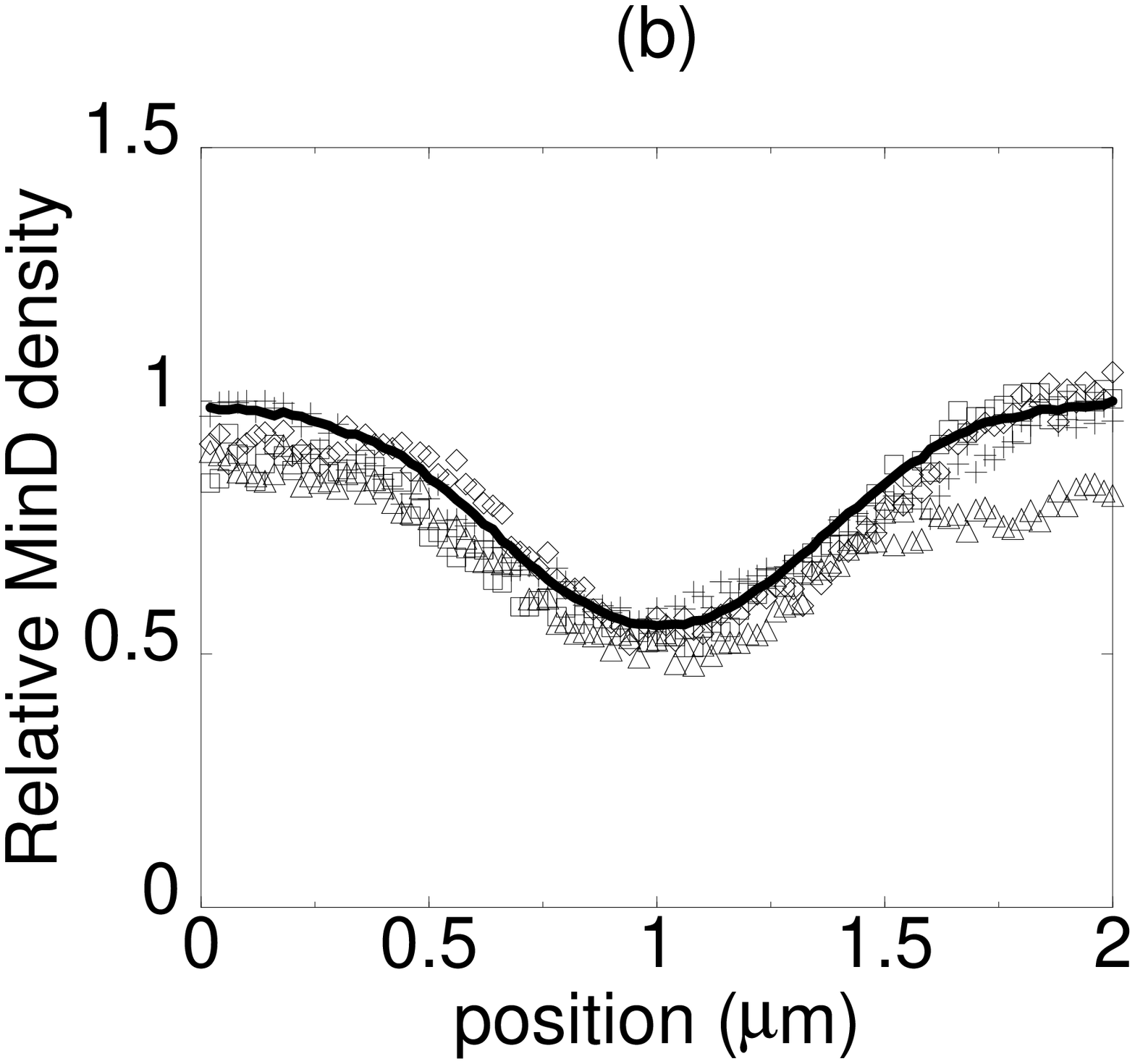}}
\vspace{.4cm}
\centerline{\hbox{
\hspace{0.2cm}
\epsfysize=1.5in
\epsfbox{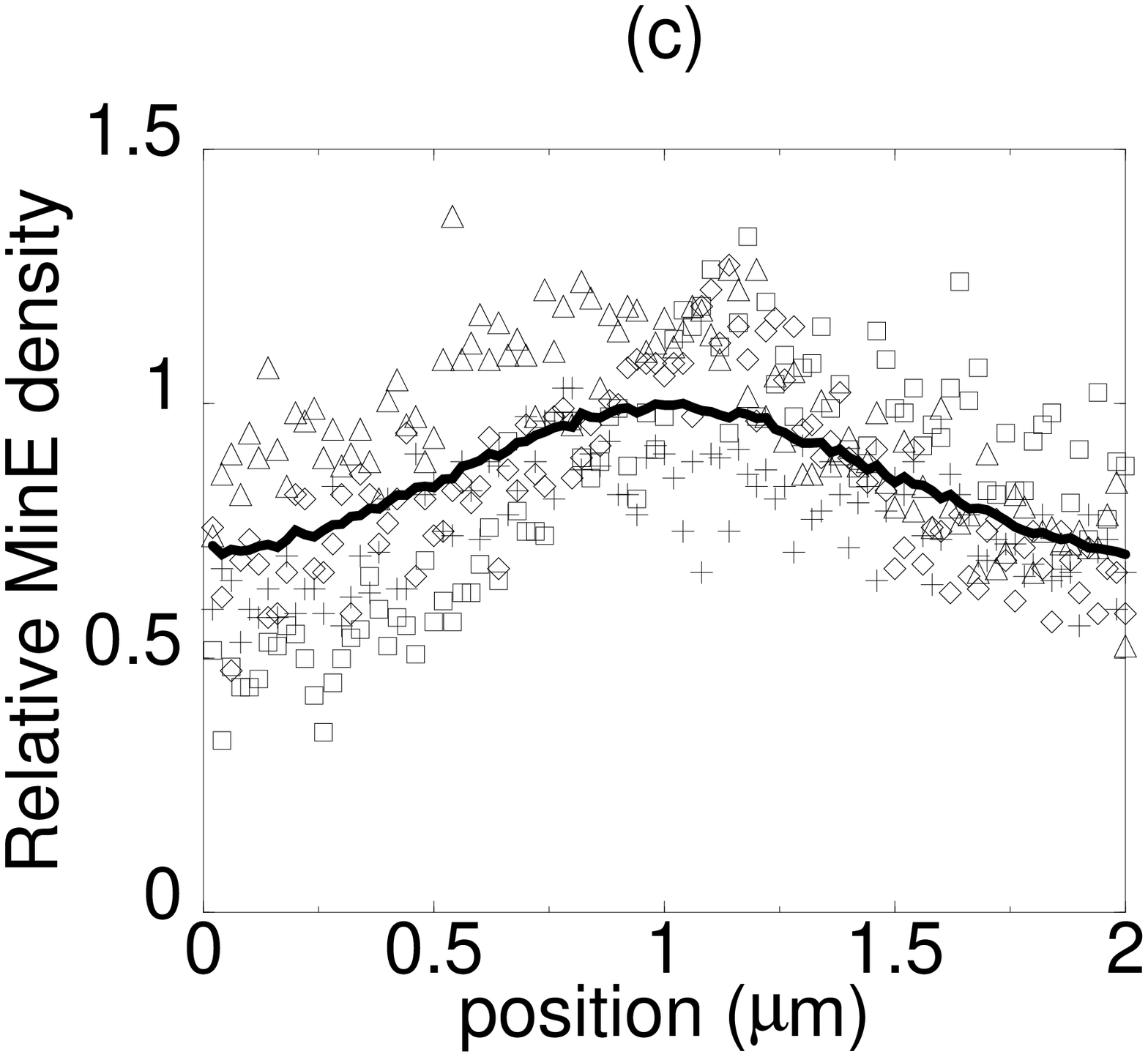}}
\hspace{0.1in}
\epsfysize=1.5in
\epsfbox{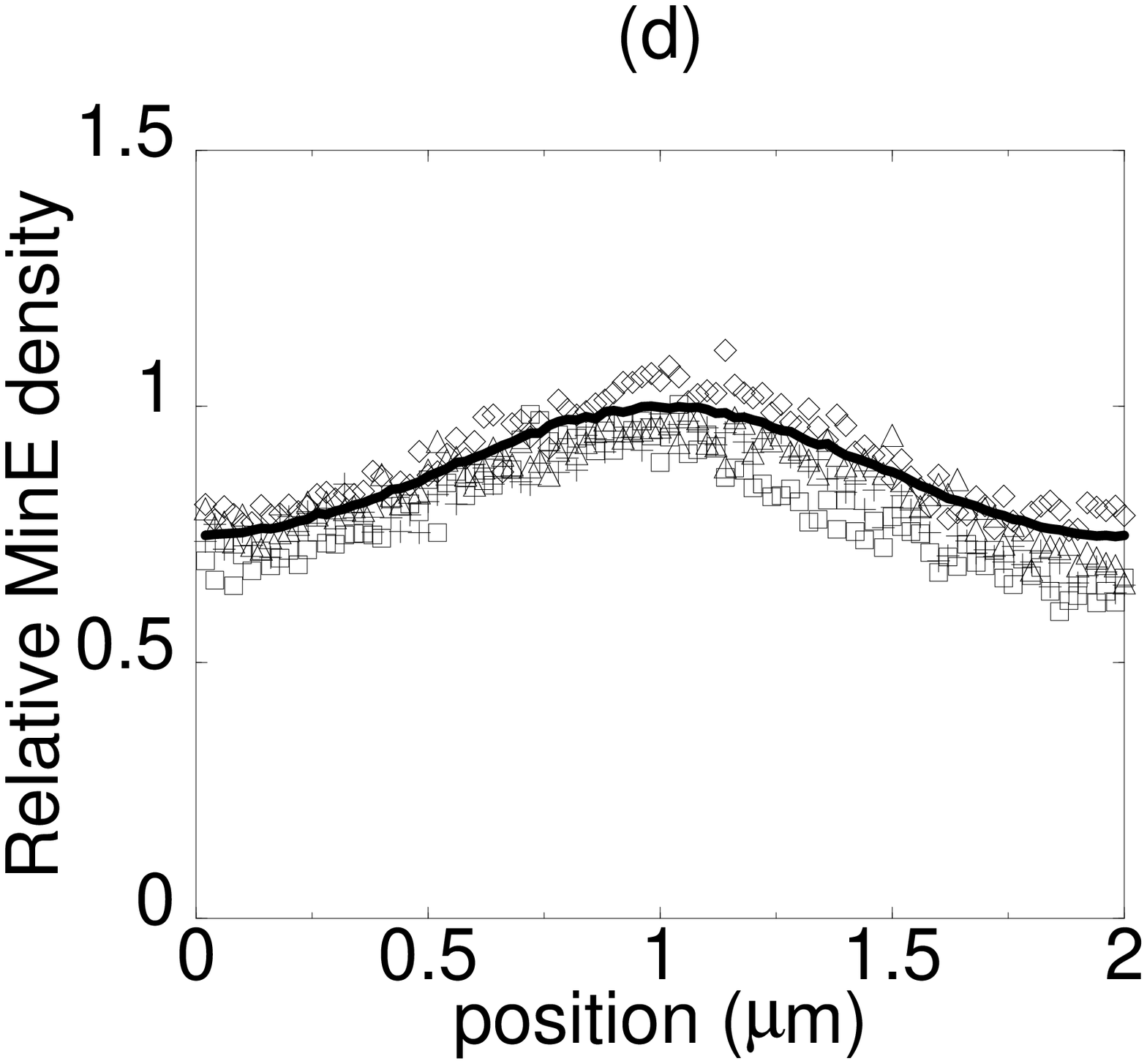}}
}
\caption{(a,b) MinD and (c,d) MinE relative density profiles in 
stochastic model with
(a,c) $N=200$ (b,d) $N=1500$. Thick lines represent averages over (a,c) 
$160$ and (b,d) $110$ consecutive cycles respectively (maximum average 
density normalized to unity). Symbols ($ +, \diamond , \bigtriangleup , 
\Box$) represent individual data sets that are averaged only over a 
single oscillation cycle.   
}
\label{avconc}
\end{figure}

{\em Comparison with experiment.}
Experimentally, the precision of the MinCDE system can be probed in
anucleate cells by measuring the position of the FtsZ ring. In these
cells the only positional guide for division is the MinCDE system, 
which functions even in the
 absence of the nucleoids. Indeed in these anucleate
cells the FtsZ ring position is placed at midcell with a width 
of $0.12~\mu m$ (scaled to a $2~\mu m$ length) \cite{Yu}, 
somewhat larger than 
the MinD distribution width $0.07~\mu m$ at $N=1500$. 
We would expect that FtsZ ring nucleation would not precisely track 
the MinCD minimum, which could account for the difference. 
In order to make this comparison we need to know how many cycles the
FtsZ nucleation averages over to identify the location of the MinD
minimum. Experimentally, this timescale is
on the order of a minute, since, from Ref.~\cite{Raskin}, oscillation
cycles of between $30~s$ and $120~s$ give normal division.
As referred to earlier, this justifies a rough comparison of the width
of the single cycle ($\sim 90~s$) MinD density minimum distribution at
$N=1500$ with that of the FtsZ distribution in anucleate cells. 

In conclusion, we have studied fluctuations in the {\em E. coli} 
MinCDE system. In some regimes, we have found that 
fluctuations are {\em essential} for pattern generation. 
We have also found evidence that the
MinCDE concentrations may be
optimal for reliable midcell division. Based on the MinCDE system, we
see that $O(10^3)$ copies of pattern forming
proteins are required in bacteria to obtain positional information
accurate to within a few percent.

\begin{figure}
\vbox{
\centerline{\hbox{
\hspace{0.0cm}
\epsfysize=1.5in
\epsfbox{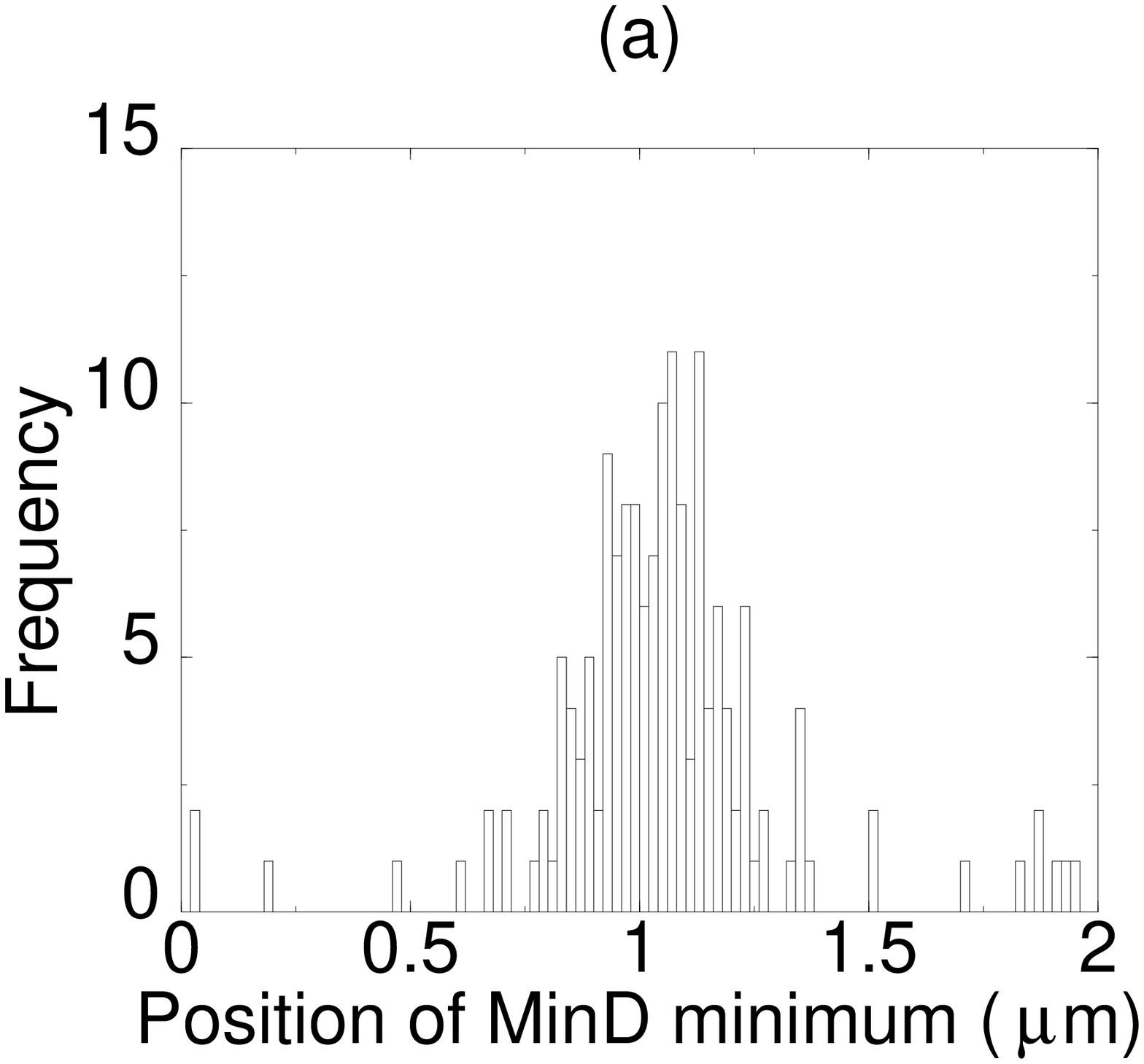}}
\hspace{0.1cm}
\epsfysize=1.5in
\epsfbox{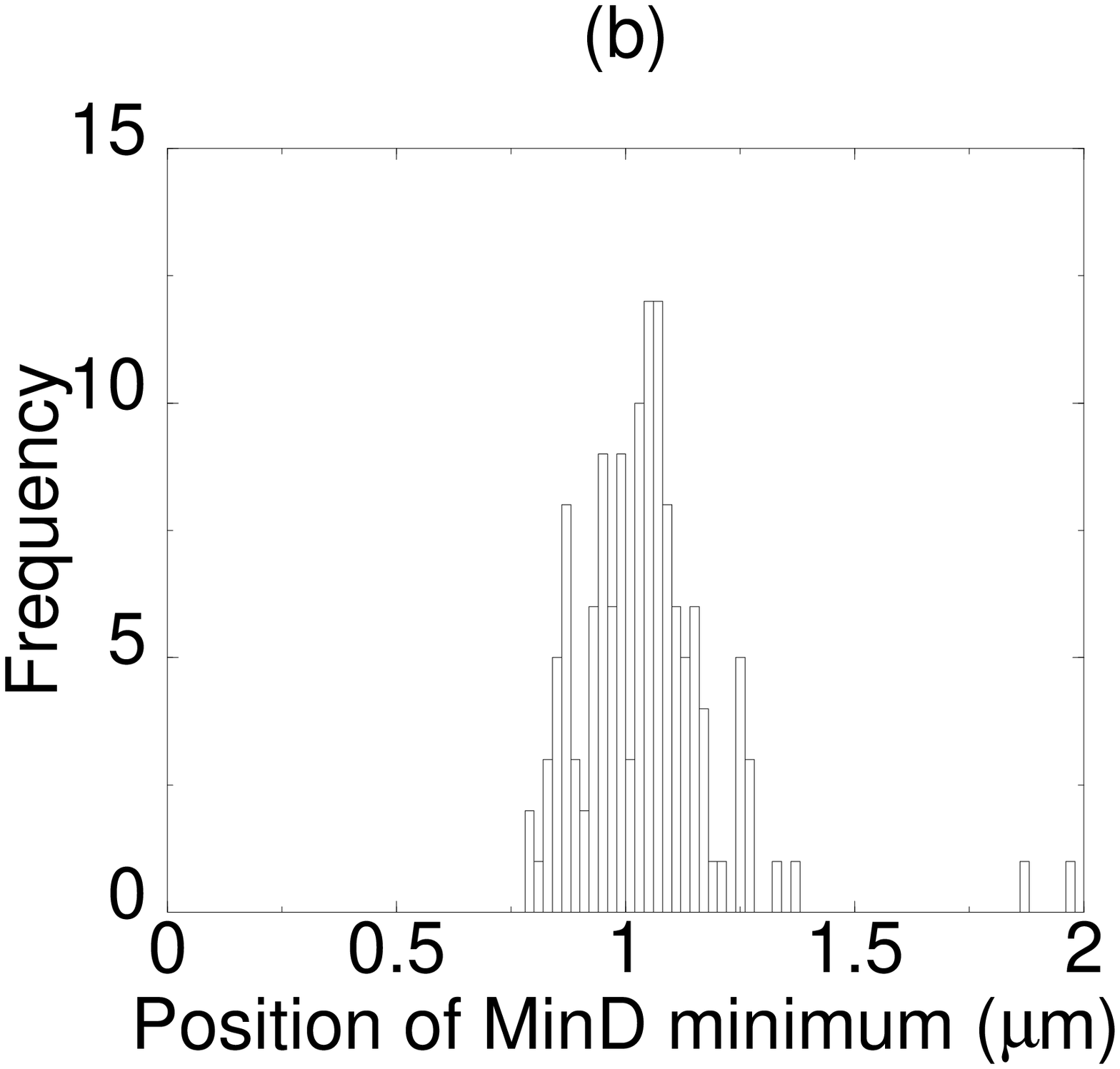}}
\vspace{.4cm}
\centerline{\hbox{
\hspace{0.0cm}
\epsfysize=1.5in
\epsfbox{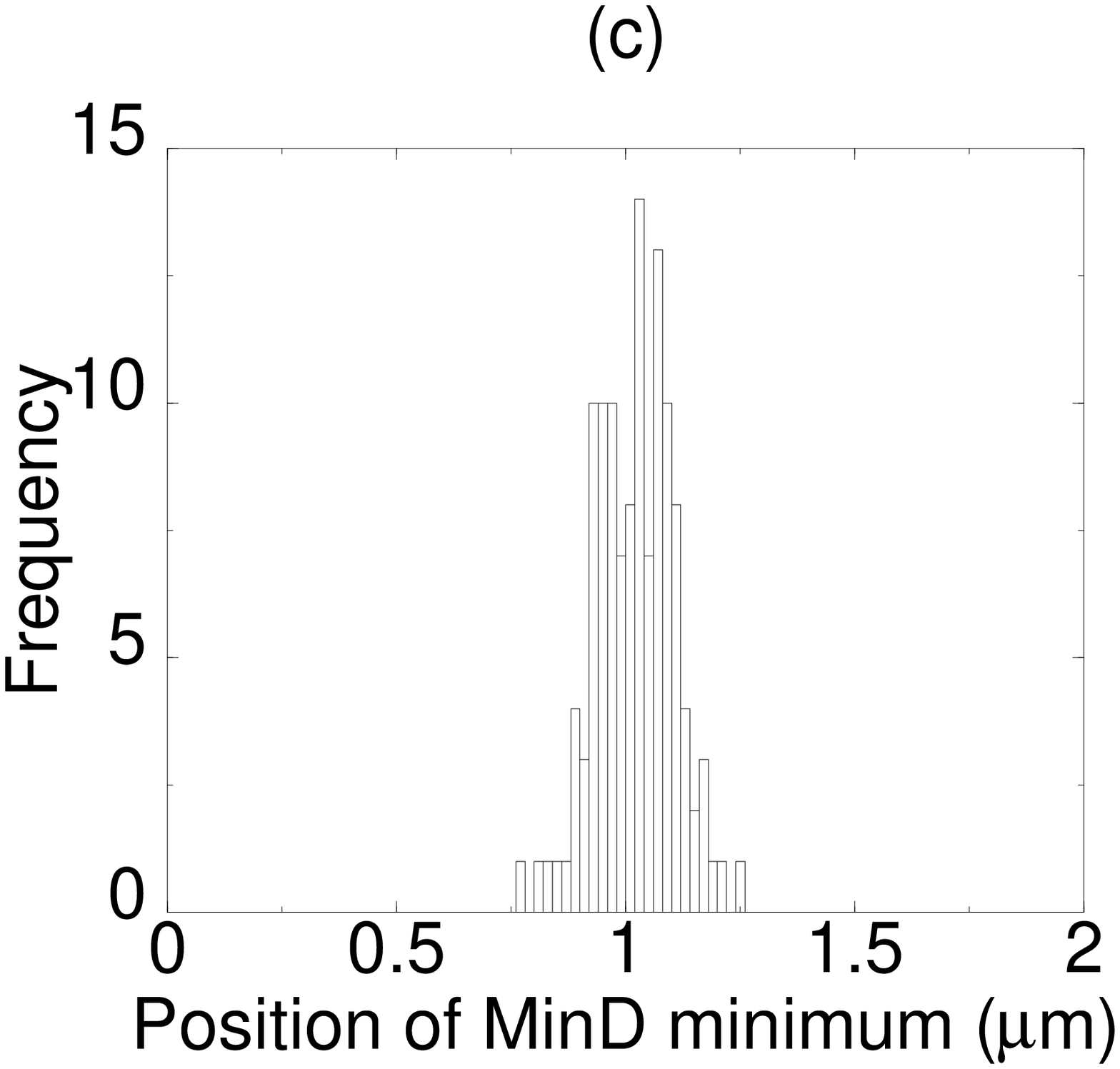}}
\hspace{0.1cm}
\epsfysize=1.5in
\epsfbox{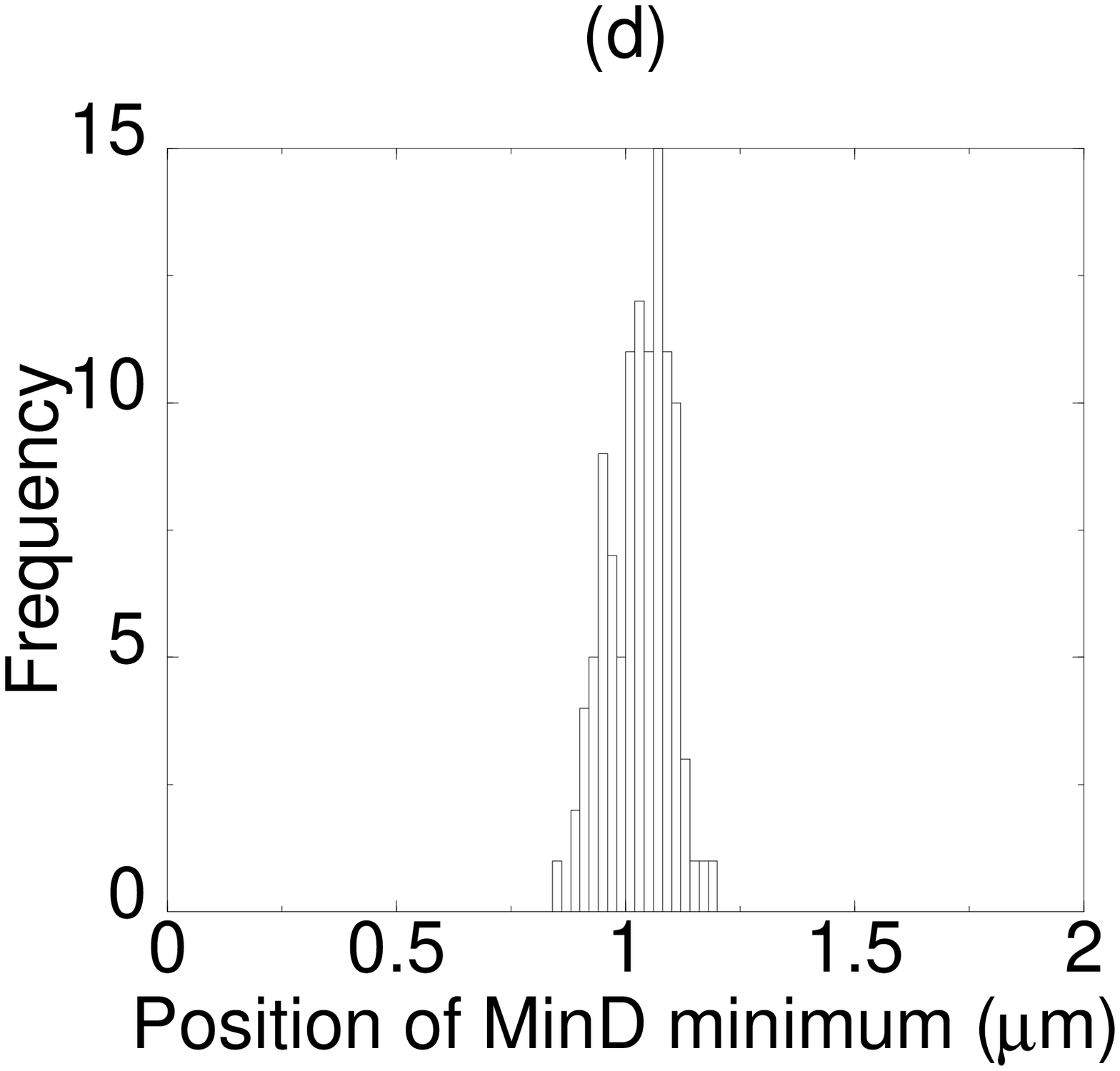}}
}
\caption{
Histogram for position of MinD density minimum in the stochastic model
with (a) $N=200$, (b) $N=400$, (c) $N=800$, (d) $N=1500$. The position
of each minimum is computed by averaging over 
a single oscillation cycle. Data is for $160$, $135$, $120$ and $110$
consecutive cycles respectively.
}
\label{centrehist}
\end{figure}

We acknowledge useful conversations with
P. de Boer, M. van Hecke, K. Kruse, J. Lutkenhaus, and
S. de Vet. 
M.H. acknowledges financial support from Stichting FOM and from The
Royal Society. A.D.R acknowledges financial support from NSERC Canada. 

\noindent
$^*$Present address: Department of Mathematics, Imperial College London,
South Kensington Campus, London SW7 2AZ, U.K.

\vspace{-0.5cm}



\begin{thebibliography}{10}

\vspace{-0.5cm}

\bibitem{Raskin} D. M. Raskin and P. A. J. de Boer,
Proc. Natl. Acad. Sci. U.S.A. {\bf 96}, 4971 (1999).

\bibitem{Raskin1} D. M. Raskin and P. A. J. de Boer, J. Bacteriol. 
{\bf 181}, 6419 (1999).

\bibitem{Hu} Z. Hu and J. Lutkenhaus, Mol. Microbiol. {\bf 34}, 82 (1999).

\bibitem{Rowland} S. L. Rowland {\em et al.}, 
J. Bacteriol. {\bf 182}, 613 (2000).

\bibitem{Fu} X. Fu {\em et al.}, 
Proc. Natl. Acad. Sci. U.S.A. {\bf 98}, 980 (2001).  

\bibitem{Rothfield} L. I. Rothfield {\em et al.}, 
Cell {\bf 106}, 13 (2001).

\bibitem{Hale} C. A. Hale {\em et al.}, 
EMBO J. {\bf 20}, 1563 (2001).

\bibitem{Yu} X.-C. Yu and W. Margolin, Mol. Microbiol. {\bf 32}, 315
(1999). 

\bibitem{Howard} M. Howard, A. D. Rutenberg, and S. de Vet,
Phys. Rev. Lett. {\bf 87}, 278102 (2001).

\bibitem{MeinhardtBoer} H. Meinhardt and P. A. J. de Boer,
Proc. Natl. Acad. Sci. U.S.A. {\bf 98}, 14202 (2001).

\bibitem{Kruse} K. Kruse, Biophys. J. {\bf 82}, 618 (2002).

\bibitem{Mcadams} H. H. McAdams and A. Arkin, Trends Genet. {\bf 15},
65 (1999).

\bibitem{Bialek} W. Bialek, in
{\it Advances in Neural Information Processing 13}, edited by
T. K. Leen, T. G. Dietterich, and  V. Tresp (MIT
Press, Cambridge, 2001), p. 103.

\bibitem{Shih} Y.-L. Shih {\em et al.}, EMBO J. {\bf 21}, 3347
(2002). 

\bibitem{Paulsson} J. Paulsson {\em et al.}, 
Proc. Natl. Acad. Sci. U.S.A. {\bf 97}, 7148 (2000).
 
\bibitem{Vilar} J. M. G. Vilar {\em et al.}, 
Proc. Natl. Acad. Sci. U.S.A. {\bf 99}, 5988 (2002). 

\bibitem{Kessler} D. A. Kessler and H. Levine, Nature {\bf 394}, 556
(1998). 

\bibitem{Falcke} M. Falcke {\em et al.}, 
Phys. Rev. E
{\bf 62}, 2636 (2000).

\bibitem{Vilar1} J. M. G. Vilar and J. M. Rub\'\i,
Phys. Rev. Lett. {\bf 78}, 2886 (1997); Physica A {\bf 277},
327 (2000).

\bibitem{Sancho} J. Garc\'\i a--Ojalvo and J. M. Sancho, {\it Noise in 
Spatially Extended Systems} (Springer--Verlag, New York, 1999).

\bibitem{Z} H. Zhonghuai {\em et al.}, Phys. Rev. Lett. {\bf 81}, 2854
(1998). 

\bibitem{deBoer} P. A. J. de Boer {\em et al.}, EMBO J. {\bf 10}, 4371
(1991). 

\bibitem{Hu02} Z. Hu {\em et al.},
Proc. Natl. Acad. Sci. U.S.A. {\bf 99}, 6761 (2002).

\bibitem{polymer} We have also simulated a stochastic $1d$ model where 
membrane MinD can polymerize and is then permitted to unbind only from 
polymer ends. Using a large membrane diffusion constant $0.1 
\mu m^2 s^{-1}$ for 
unpolymerized membrane MinD, zero diffusion for polymerized MinD, and
otherwise unchanged parameters, we again recover 
oscillatory dynamics, with an average midcell MinD concentration 
minimum. Thus, the key feature introduced by the polymerization 
is the small membrane diffusivity.

\bibitem{param}
As $N$ is scaled, so must the reaction rates in order to maintain the   
same equivalent deterministic model.  This ensures that 
the variation in widths is primarily due 
to changes in fluctuations (induced by varying $N$), and
not to changes in the membrane binding/unbinding rates. 

\bibitem{sun1998} Q. Sun {\em et al.}, Mol. Microbiol. {\bf 29}, 491
(1998). 
  
\end{thebibliography}
\end{document}